\documentclass[12pt]{article}
\usepackage{graphicx}

\def\napoli{P-25, MS H846, Physics Divison, LANL, Los Alamos, NM, 87545, USA}
\def\support{\footnote{email address: xuanlipx@rcf.rhic.bnl.gov}}

\def\Title#1{\begin{center} {\Large #1 } \end{center}}
\def\Author#1{\begin{center}{ \sc #1} \end{center}}
\def\Address#1{\begin{center}{ \it #1} \end{center}}

\newenvironment{Abstract}{\begin{quotation}  }{\end{quotation}}
\newenvironment{Presented}{\begin{quotation} \begin{center} 
             PRESENTED AT\end{center}\bigskip 
      \begin{center}\begin{large}}{\end{large}\end{center} \end{quotation}}

\def\beq{\begin{equation}}
\def\eeq#1{\label{#1}\end{equation}}
\def\eeqn{\end{equation}}

\def\beqa{\begin{eqnarray}}
\def\eeqa#1{\label{#1}\end{eqnarray}}
\def\eeqan{\end{eqnarray}}

\let\bar=\overbar

\def\Dslash{\not{\hbox{\kern-4pt $D$}}}
\def\dslash{\not{\hbox{\kern-2pt $\del$}}}

\def\msb{{\bar{\ssstyle M \kern -1pt S}}}

\begin{document}
\begin{titlepage}

\vfill
\Title{Energy and system size dependent heavy flavor measurements at PHENIX at RHIC}
\vfill
\Author{ Xuan Li for the PHENIX collaboration \support}
\Address{\napoli}
\vfill
\begin{Abstract}
Heavy flavor production is an ideal tool to study the properties of the Quark Gluon Plasma (QGP). The heavy flavor production at the Relativistic Heavy Ion Collider (RHIC) has its unique kinematic coverage and different production mechanisms from the Larger Hadron Collider (LHC) measurements. Heavy flavor products created in heavy ion collisions experience the whole evolution of nuclear medium. It's critical to measure both open and hidden heavy flavor products in different collision systems to isolate cold/hot nuclear medium effects and initial/final state interactions. We report recent heavy flavor measurements at PHENIX in 200 GeV $p$+$p$, $p$+Al, $p$+Au, $^{3}$He+Au and Au+Au collisions that include: the correlated di-muon analysis in 200 GeV $p$+$p$ and $p$+Au collisions; the rapidity and $N_{part}$ dependent $J/\psi$ $R_{AB}$ measured in asymmetric small systems; open heavy flavor $v_{2}$ measured in 200 GeV d+Au and Au+Au collisions. These measurements provide further information about the heavy flavor production mechanism, initial and final state nuclear modification and flavor dependent energy loss in QGP.
\end{Abstract}
\vfill
\begin{Presented}
Thirteenth Conference on the Intersections of Particle and Nuclear Physics, \\
May 29th to June 3th, 2018 - Palm Springs, CA, USA.
\end{Presented}
\vfill
\end{titlepage}
\def\thefootnote{\fnsymbol{footnote}}
\setcounter{footnote}{0}

\section{Introduction}
The Heavy quark production has been used to test the Quantum Chromodynamics (QCD) calculations. In addition to this, the heavy flavor production is one of the ideal hard probes to study the whole evolution of the Quark Gluon Plasma (QGP) as it is produced earlier than the QGP formation due to its high mass ($m_{c}$/$m_{b} >> \Lambda_{QCD}$). The PHENIX experiment at the Relativistic Heavy Ion Collider (RHIC) have collected high luminosity data sets for both open and hidden heavy flavor measurements in 200/510 GeV $p$+$p$, 200 GeV $p$+Al, $p$+Au, d+Au, $^{3}$He+Au, Cu+Au, Cu+Cu, Au+Au and U+U collisions. We summarize recent heavy flavor measurements at PHENIX including: the dimuon correlation measurements within $1.2<|y|<2.2$ in 200 GeV $p$+$p$ and $p$+Au collisions to understand the charm and bottom production mechanisms and the cold nuclear modification of open heavy production; the inclusive $J/\psi$ nuclear modification measurements in 200 GeV $p$+Al, $p$+Au and $^{3}$He+Au collisions that help disentangle the initial and final state effects on the charmonium; the elliptic flow measurements of open heavy flavor production via semileptonic decayed single muons at forward rapidity in 200 GeV d+Au collisions and through semileptonic decayed single electrons at mid-rapidity in 200 GeV Au+Au collisions. These measurements provide new clues about how heavy flavor production gets modified in cold and hot nuclear medium and shed light on future measurements.

\section{Correlated dimuon measurements in $p$+$p$ and $p$+Au collisions}
\begin{figure}[!ht]
\centering
	\includegraphics[width=0.42\textwidth]{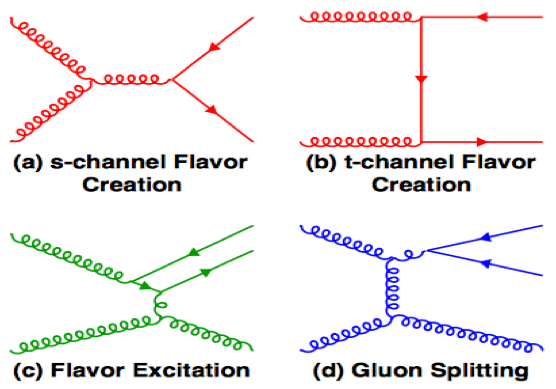}
\caption{Partonic processes for heavy flavor production: flavor creation (a, b), flavor excitation (c) and gluon splitting (d).}
\label{fig:hf_prod}
\end{figure}

Perturbative QCD calculations \cite{hf_prod} predict that heavy quarks are generated in hadron collisions through flavor creation, flavor excitation and gluon splitting processes as shown in Figure~\ref{fig:hf_prod}. The heavy flavor production mechanism is center of mass dependent. To understand the production processes of charm and bottom quarks at RHIC energy, the like-sign and unlike-sign dimuon pairs were analyzed within $1.2<|y|<2.2$ rapidity region in 2015 200 GeV $p$+$p$ data at PHENIX. 

\begin{figure}[!ht]
\centering
\includegraphics[width=0.38\linewidth]{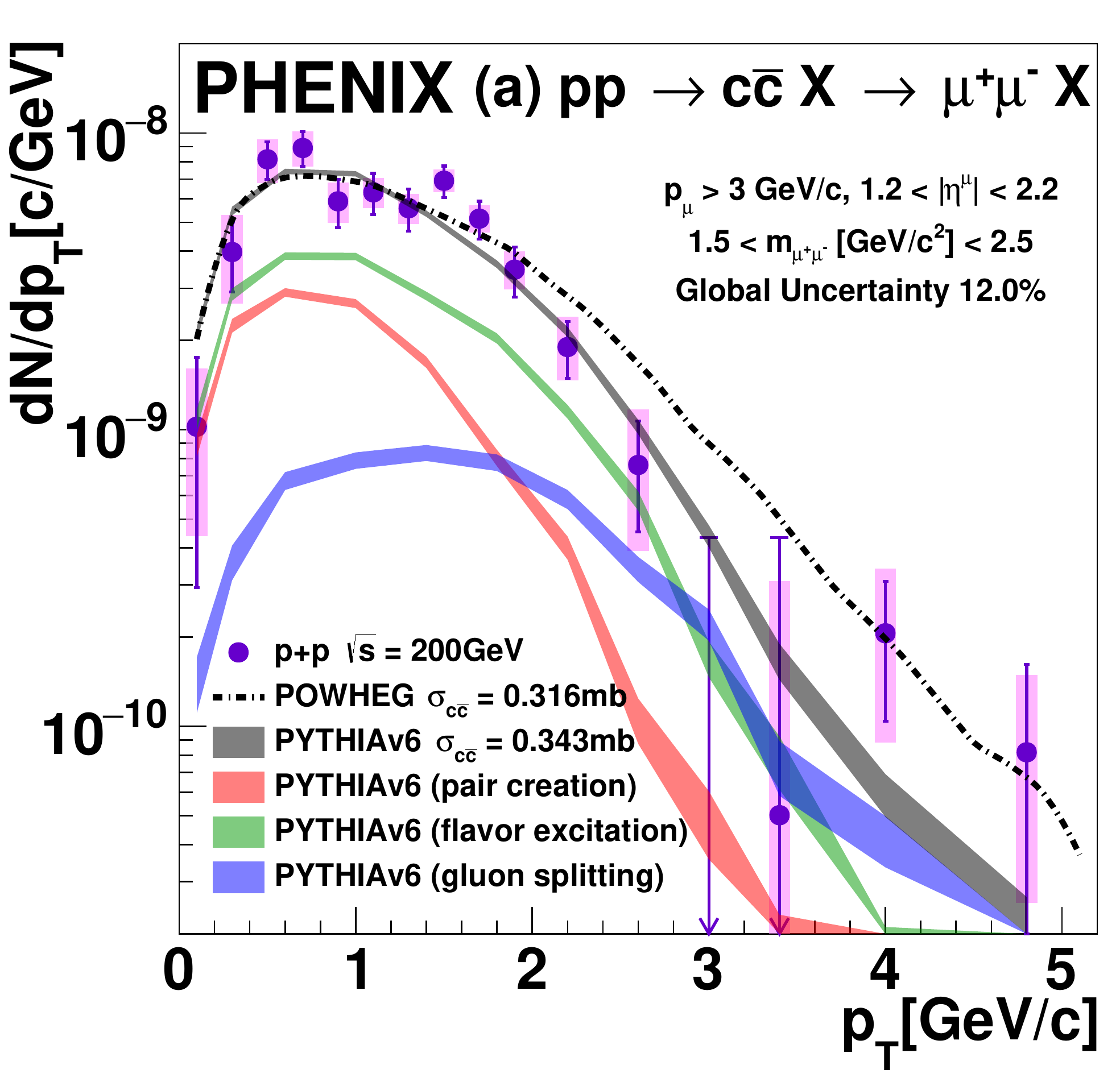}
\includegraphics[width=0.38\linewidth]{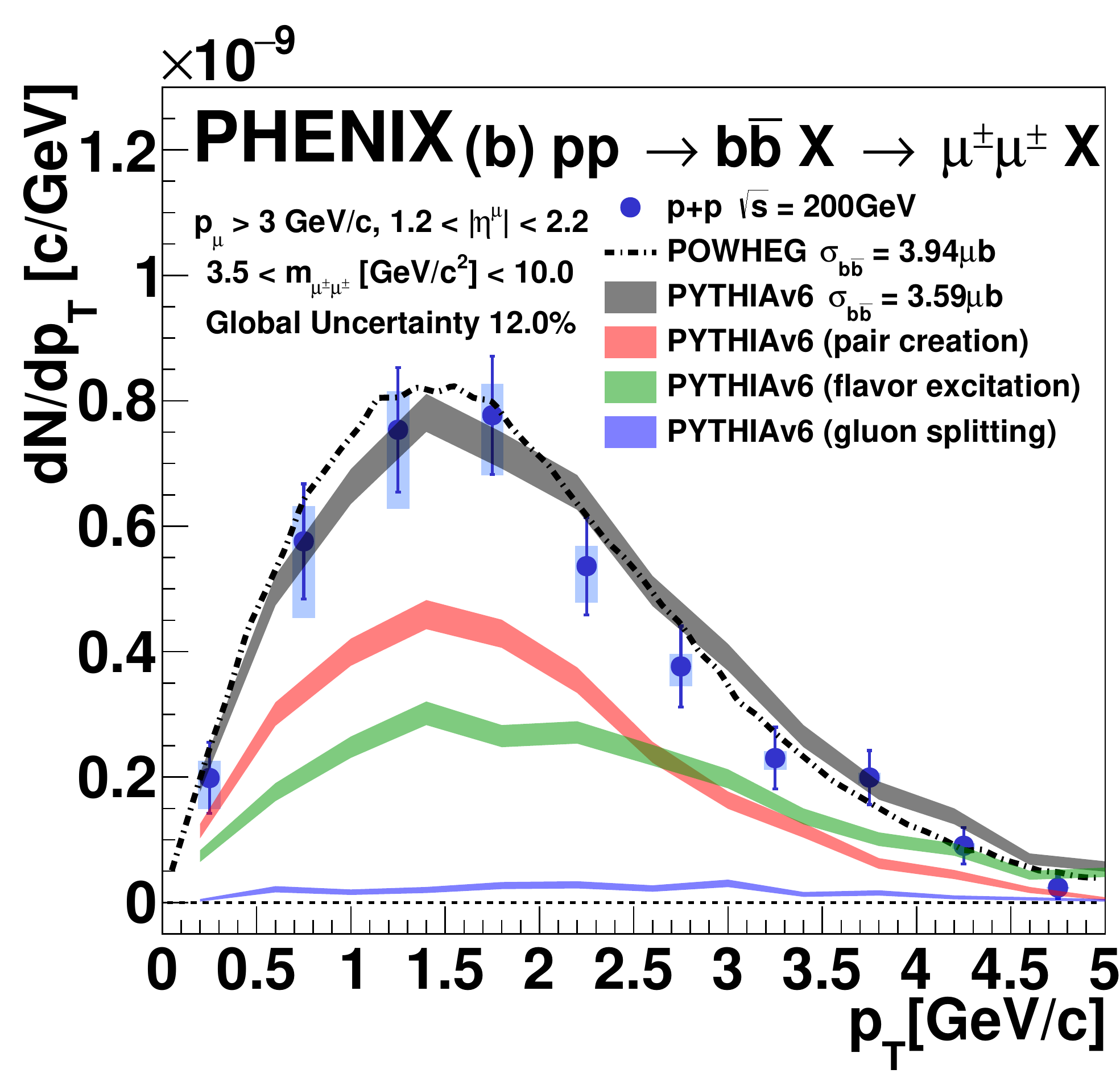}
\caption{\label{fig:dimu_pp} The corrected dimuon yield from charm (left) and bottom (right) decays versus the muon pair $p_{T}$. Data are compared to distributions calculated with PYTHIA and POWHEG that normalized to data. The generated distributions of the dimuon pair yields from PYTHIA are further divided into subsamples from pair creation (flavor creation), flavor excitation, and gluon splitting partonic processes.}
\end{figure}

\begin{figure}[!ht]
\centering
	\includegraphics[width=0.4\textwidth]{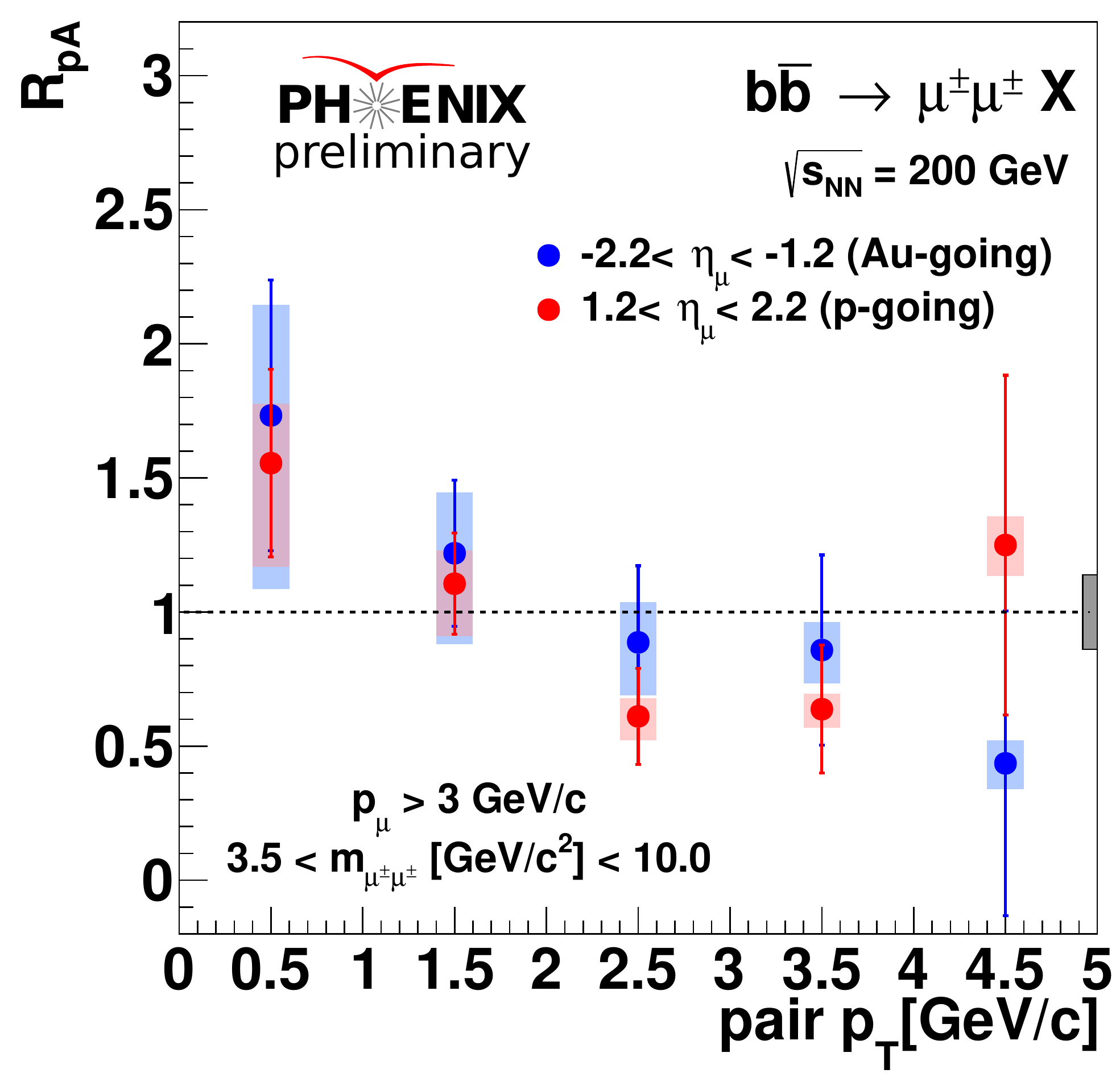}
\caption{R$_{pAu}$ of bottom decayed dimuons versus the dimuon pair $p_{T}$ in 200 GeV $p$+Au collisions. The correlated dimuon yields from bottom decays are determined in the like-sign 3.5 GeV/$c^{2} < m_{\mu{\pm}}m_{\mu{\pm}} <$ 10 GeV/$c^{2}$ mass region.}
\label{fig:dimu_pAu}
\end{figure}

To reduce the hadronic and jet background contribution, the muon track $p_{T}$ is required to be large than 3 $GeV/c$. With extrapolated hadronic cocktail background, the dimuon pair from $b\bar{b}$ is determined through a simultaneous fit in high mass (3.5 GeV/$c^{2} < m_{\mu{\pm}}m_{\mu{\pm}} <$ 10 GeV/$c^{2}$) like-sign dimuon mass and $p_{T}$ spectra. The charm ($c\bar{c}$) contribution is determined from unlike-sign di-muon pairs and has the highest signal to background fraction within 1.5 GeV/$c^{2} < m_{\mu{\pm}}m_{\mu{\mp}} <$ 2.5 GeV/$c^{2}$ mass region. The spectra of azimuthal open angle correlation and the muon pair $p_{T}$ from $b\bar{b}$ and $c\bar{c}$ measured in data are compared with distributions generated in PYTHIA \cite{pythia} and POWHEG \cite{powheg}, which both include the next-to-leading order processes. As shown in Figure~\ref{fig:dimu_pp}, distributions of dimuon pair $p_{T}$ from charm and bottom decays are in reasonable agreements with PYTHIA calculations. The POWHEG calculation is consistent with data for dimuon pairs with $p_{T} < $ 2 GeV$/c$ but is significantly higher than data in the $p_{T} >$ 2 GeV$/c$ region. To evaluate the relevant partonic subprocess contribution to the heavy flavor production, the $p_{T}$ distributions of correlated dimuons that from charm or bottom decays are generated for each partonic interaction process in PYTHIA. The fraction of flavor creation, flavor excitation and gluon splitting processes are determined through the maximum log-likelihood fit to data. This study suggests the dominated production process for charm within $1.2<|y|<2.2$ in 200 GeV $p$+$p$ collisions is flavor excitation, and the bottom production in the same kinematic region is dominated by the flavor creation (pair creation) process.

Similar analysis procedure has been applied in 2015 200 GeV $p$+Au data to extract yields of correlated dimuons from bottom decays in the 3.5 GeV/$c^{2} < m_{\mu{\pm}}m_{\mu{\pm}} <$ 10 GeV/$c^{2}$ mass region. The nuclear modification factor R$_{pAu}$ of bottom decayed dimuons are measured versus the dimuon pair $p_{T}$ as shown in Figure \ref{fig:dimu_pAu}. No significant nuclear modification of bottom decayed dimuons with pair $p_{T}$ from 0 to 5 GeV$/c$ in both $p$-going and Au-going directions. This preliminary result is consistent with the binary scaling of bottom quarks in 200 GeV $p$+Au collisions.

\section{$J/\psi$ $R_{AB}$ measurements in small systems}
Rapidity dependent relative ratio of $\psi \prime$ to $J/\psi$ has been measured via dimuon channel within $1.2<|y|<2.2$ at PHENIX in 200 GeV $p$+Al, $p$+Au and $^{3}$He+Au collisions \cite{psi_dratio}. In the light nuclei going direction, no significant relative suppression of $\psi \prime$ to $J/\psi$ has been measured. While in the heavy nuclei direction, the $\psi \prime$ yields are more suppressed than the $J/\psi$ yields. As $\psi \prime$s and $J/\psi$s are expected to experience the same initial state effects, the suppression of $\psi \prime$ to $J/\psi$ double ratio indicates non-negligible final state interactions on the $\psi \prime$ production in small systems. To study the cold nuclear medium effects on the $J/\psi$ production, the rapidity and N$_{part}$ dependent nuclear modification factor R$_{AB}$ of $J/\psi$ has been measured within $1.2<|y|<2.2$ at PHENIX in 200 GeV $p$+Al, $p$+Au and $^{3}$He+Au collisions. 

\begin{figure}[!ht]
\centering
	\includegraphics[width=0.32\textwidth]{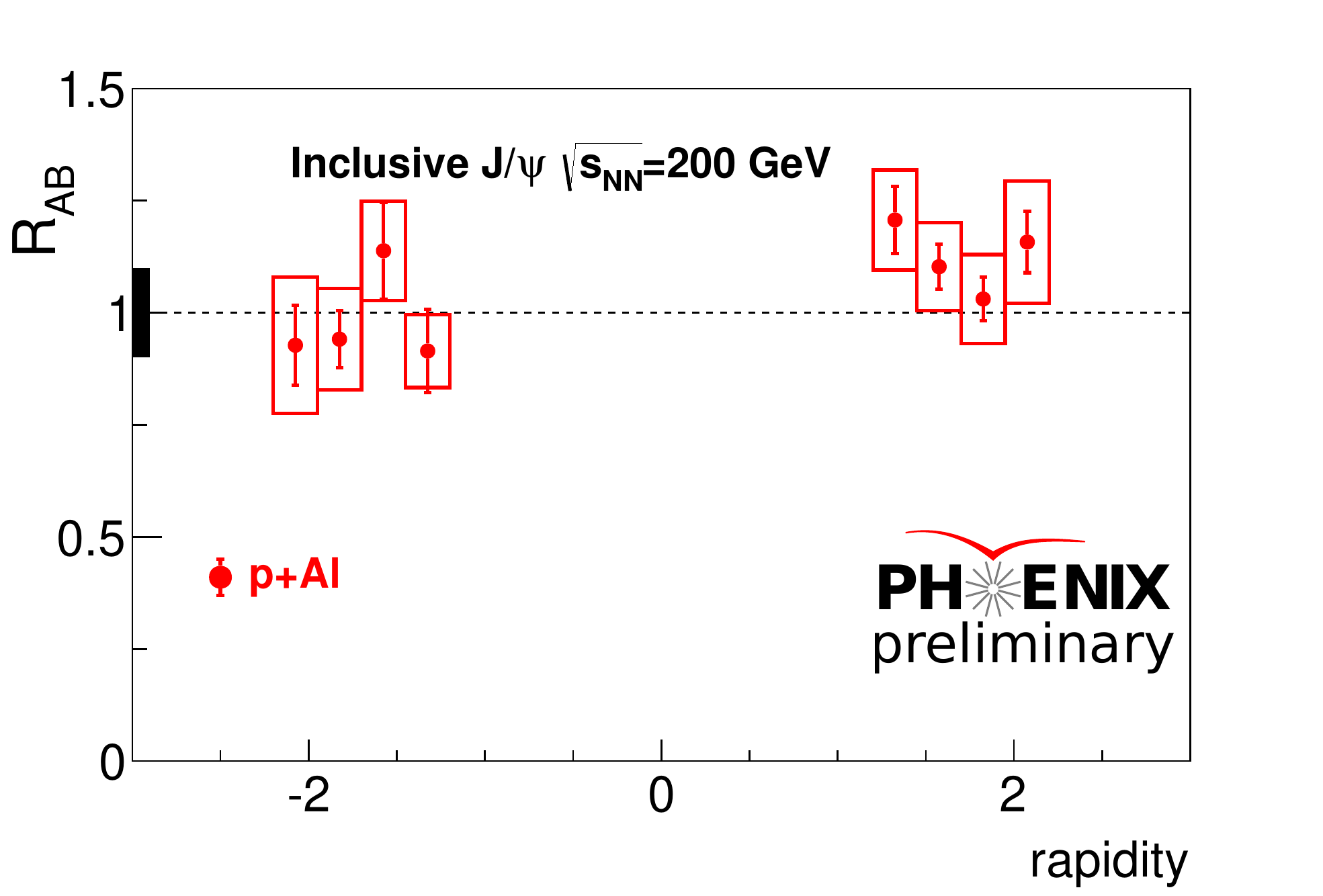}
        \includegraphics[width=0.32\textwidth]{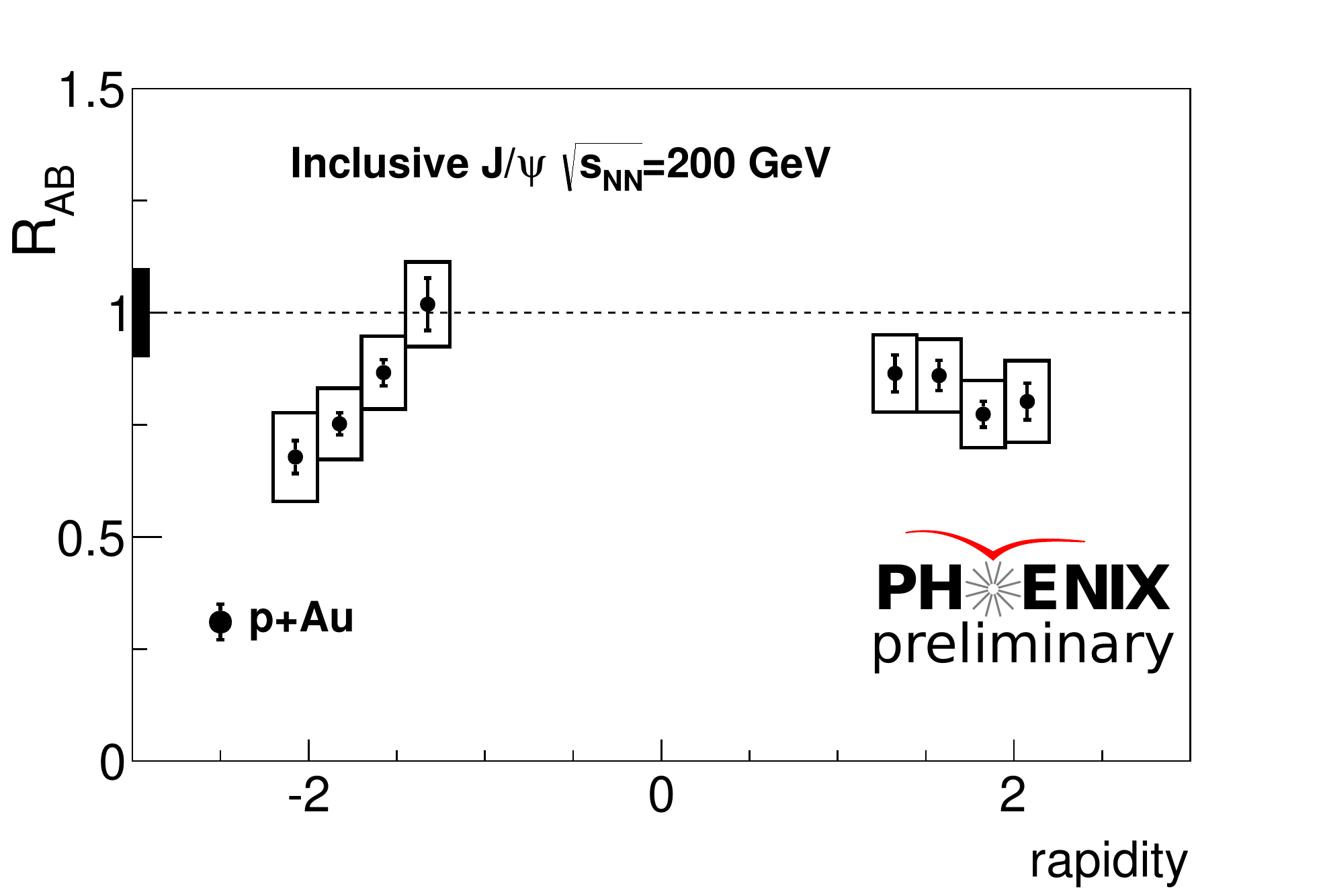}
        \includegraphics[width=0.32\textwidth]{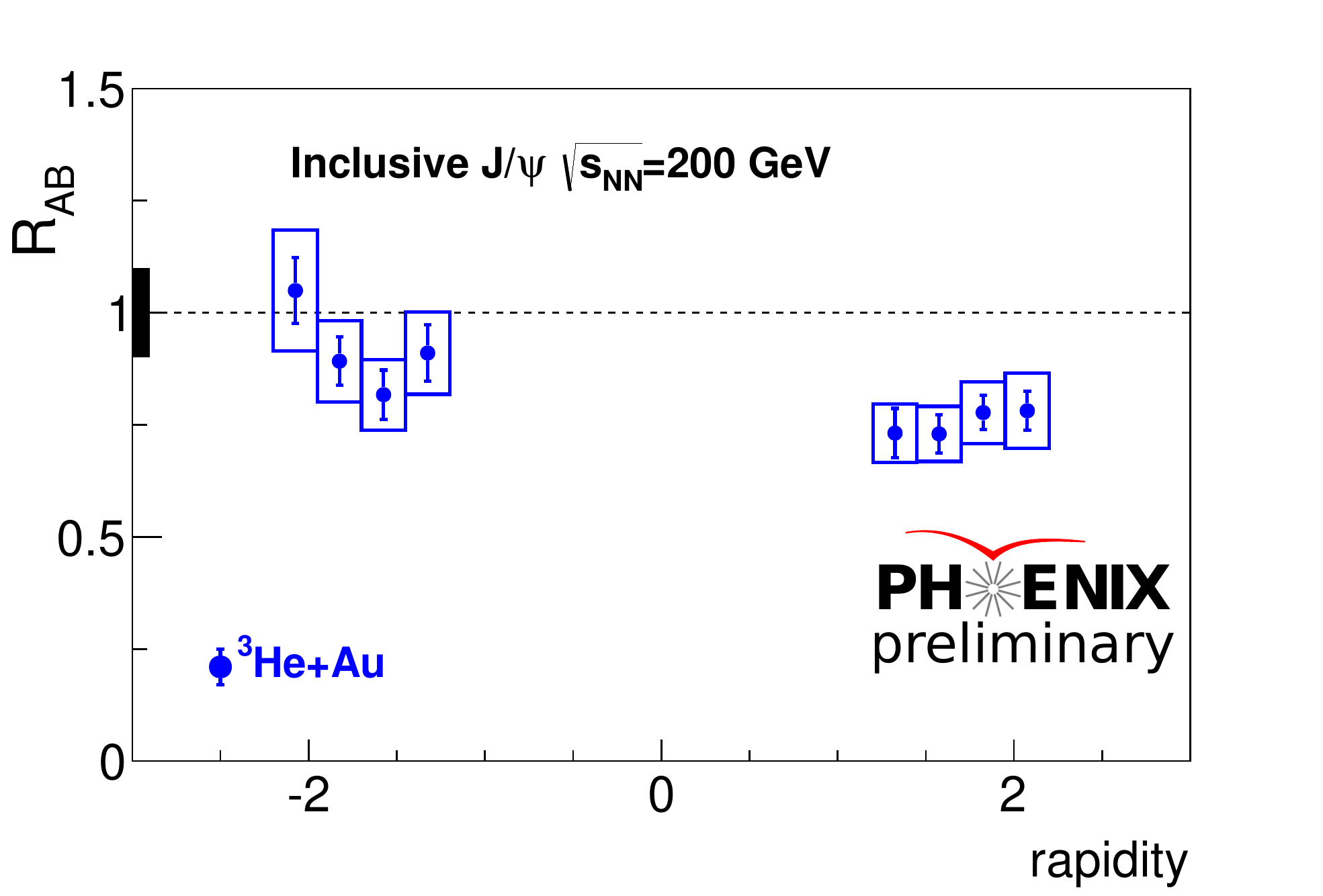}
\caption{Preliminary results of the nuclear modification factor of inclusive $J/\psi$ versus rapidity in 200 GeV $p$+Al (left), $p$+Au (middle) and $^{3}$He+Au (right) collisions. Global scale uncertainty is shown in the left solid box.}
\label{fig:jpsi_rapidity}
\end{figure}

\begin{figure}[!ht]
\centering
	\includegraphics[width=0.42\textwidth]{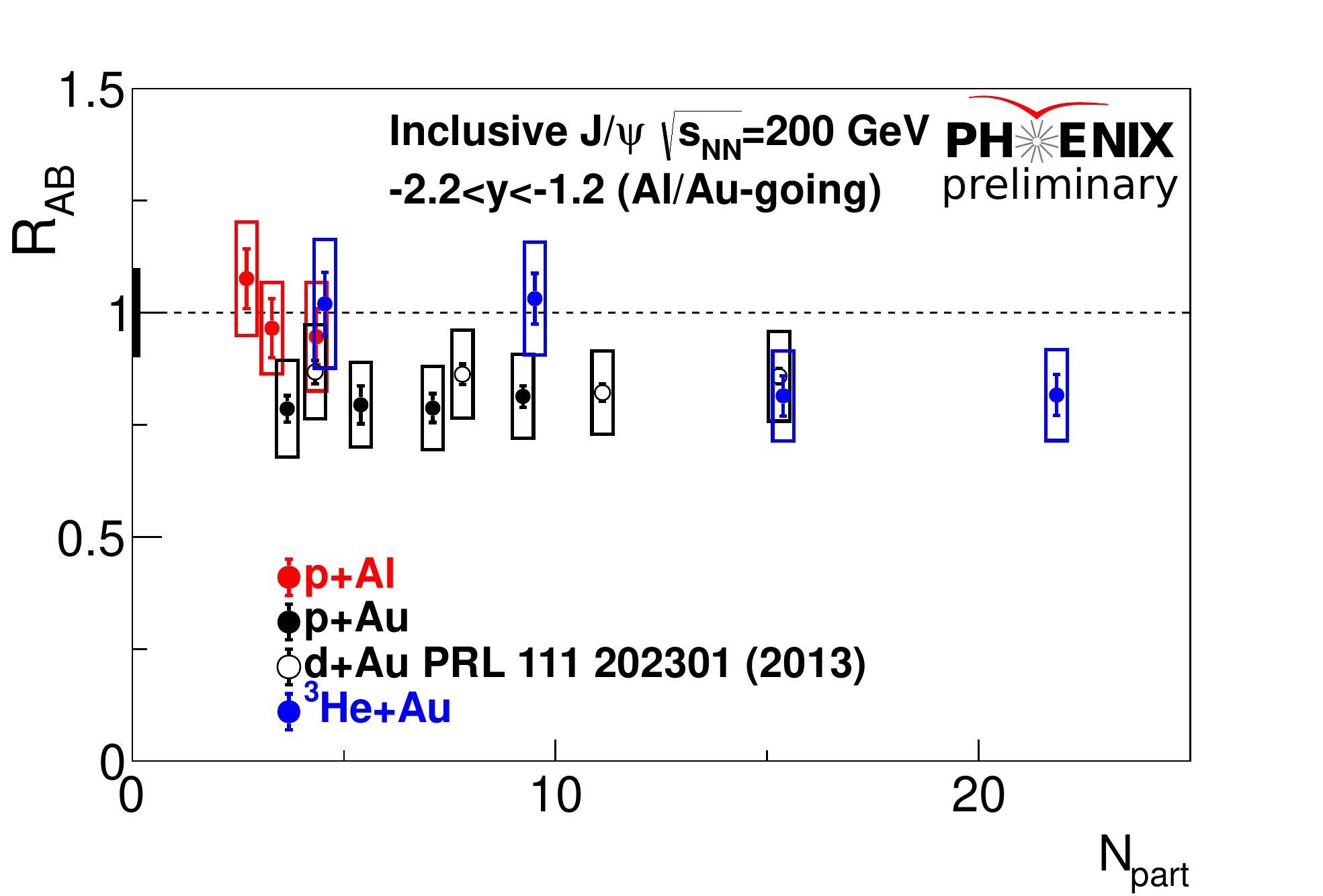}
        \includegraphics[width=0.42\textwidth]{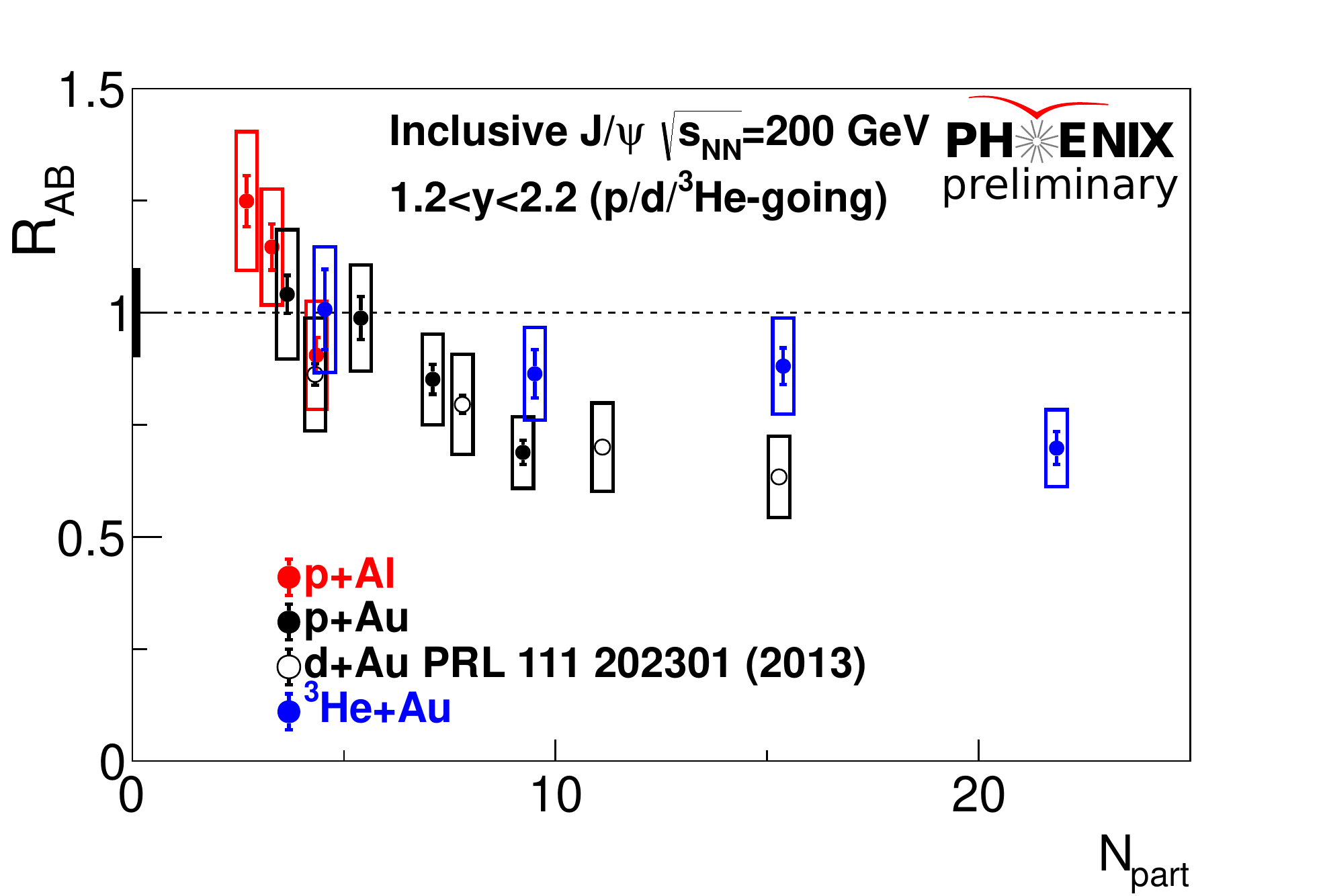}
\caption{Preliminary results of the nuclear modification factor of inclusive $J/\psi$ versus N$_{part}$ within $-2.2<y<-1.2$ rapidity region (left) and within $1.2<y<2.2$ rapidity region (right) in 200 GeV $p$+Al, $p$+Au and $^{3}$He+Au collisions. The $N_{part}$ dependent mid-rapidity $J/\psi$ nuclear modification factor results measured in 200 GeV d+Au collisions are included for reference. Global scale uncertainty is shown in the left solid box.}
\label{fig:jpsi_ncoll}
\end{figure}

Figure~\ref{fig:jpsi_rapidity} shows the rapidity dependent nuclear modification factor R$_{AB}$ of inclusive $J/\psi$ within $1.2<|y|<2.2$ in 200 GeV $p$+Al, $p$+Au and $^{3}$He+Au collisions. At forward rapidity which is the light nuclei going direction, the inclusive $J/\psi$ yields are suppressed in $p$+Au and $^{3}$He+Au collisions but not suppressed in $p$+Al collisions. At backward rapidity which is the heavy nuclei going direction, no significant nuclear modification of inclusive $J/\psi$ has been observed. The slightly system size dependent suppression of inclusive $J/\psi$ R$_{AB}$ results measured at forward rapidity suggests the nuclear dependent initial state effects dominate this region. The number of participants $N_{part}$ dependent inclusive $J/\psi$ R$_{AB}$ measured at forward and backward rapidities in 200 GeV $p$+Al, $p$+Au and $^{3}$He+Au collisions are shown in Figure \ref{fig:jpsi_ncoll}. Similar $J/\psi$ R$_{AB}$ values have been measured at a similar $N_{part}$ no matter what the beam species are. In both forward and backward rapidity regions, the $J/\psi$ R$_{AB}$ get more suppressed as $N_{part}$ increases. These results suggest both initial and final state effects contribute to the inclusive $J/\psi$ nuclear modification.

\section{Open heavy flavor $v_{2}$ measurements in d+Au and Au+Au collisions}
\begin{figure}[!ht]
\centering
	\includegraphics[width=0.7\textwidth]{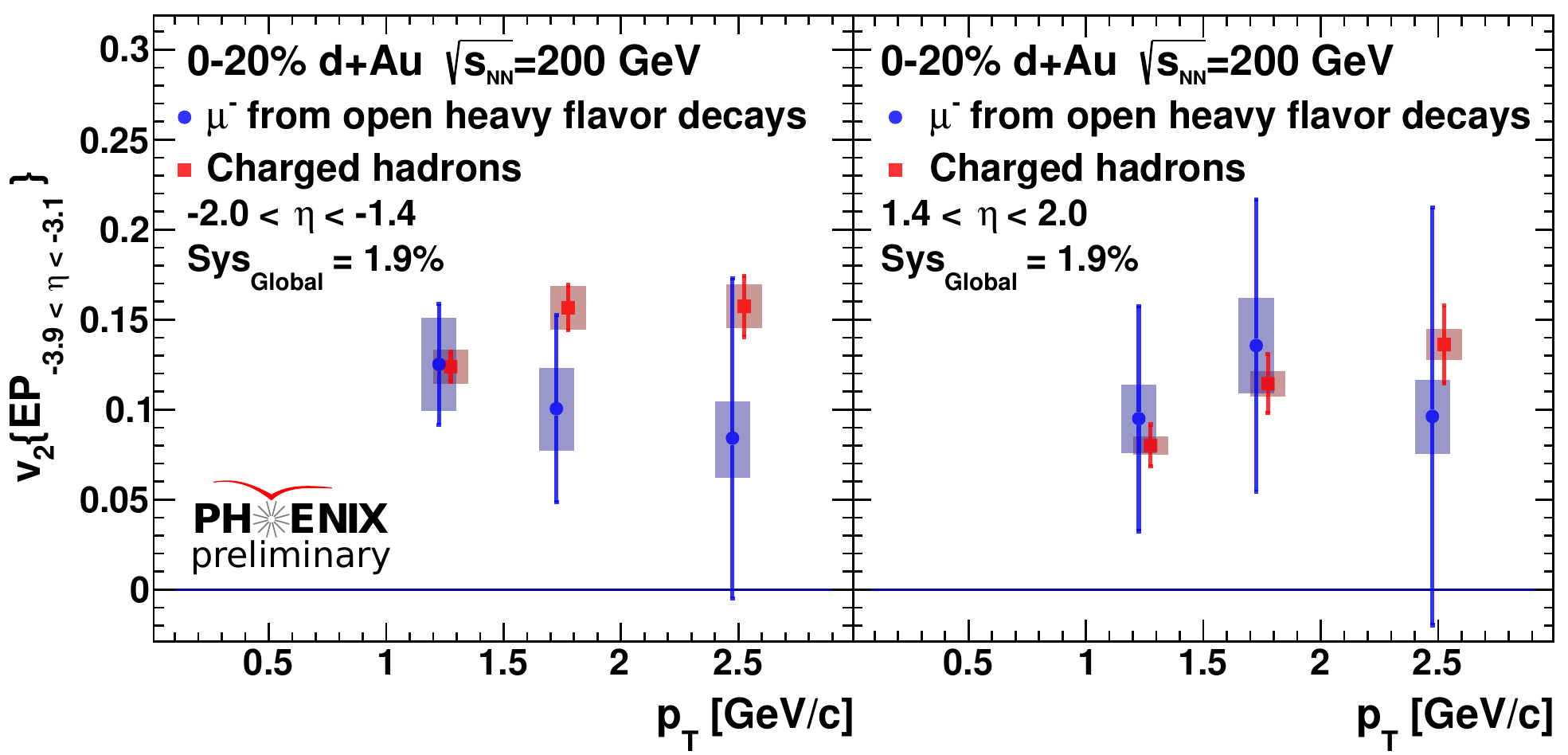}
\caption{Heavy flavor semi-leptonic decayed single $\mu^{-}$ v$_{2}$ versus $p_{T}$ within $1.4<|\eta|<2.0$ region in 0-20$\%$ d+Au collisions at $\sqrt{s} =$ 200 GeV. The heavy flavor decayed $\mu^{-}$ v$_{2}$ (blue) within $-2.0<\eta<-1.4$ ($1.4<\eta<2.0$) is shown in the left (right) panel in comparison with charged hadron v$_{2}$ (red). The global scaling systematic uncertainty is 1.9$\%$.}
\label{fig:hfmu_v2}
\end{figure}

Significant azimuthal anisotropy of light and heavy quark products have been observed at LHC in 5.02 TeV \cite{lhc_had_v2, alice_jpsi_v2} and 8.16 TeV \cite{cms_pPb} $p$+Pb collisions. The PHENIX experiment also measured significant azimuthal anisotropy of light hadrons in 200 GeV central $p$+Au, d+Au and $^{3}$He+Au collisions \cite{had_v2}. It is important to verify whether heavy quarks flow in small systems at RHIC energy. With the reaction plane method, PHENIX released the first preliminary result of the heavy flavor semi-leptonic decayed single $\mu^{-}$ v$_{2}$ within $1.4<|\eta|<2.0$ pseudorapidity region in 0-20$\%$ d+Au collisions at $\sqrt{s} =$ 200 GeV (see Figure~\ref{fig:hfmu_v2}). 

Non-zero v$_{2}$ of the heavy flavor semi-leptonic decayed $\mu^{-}$ with $p_{T}<2$ GeV$/c$ has been observed at both forward and backward pseudorapidities in central d+Au collisions. The magnitude of the heavy flavor decayed $\mu^{-}$ v$_{2}$ is comparable to the charged hadron v$_{2}$ in the low $p_{T}$ region. As the heavy flavor production at RHIC evolves the gluon fusion and flavor excitation processes, the final state heavy flavor v$_{2}$ may not only come from the heavy flavor flow but also the quark/gluon flow. In addition to initial geometry effects, a small QGP droplet formation in small system might be another explaination for the heavy flavor flow. Further theory calculations are required to help understand the heavy flavor flow in central d+Au collisions.

\begin{figure}[!ht]
\centering
	\includegraphics[width=0.45\textwidth]{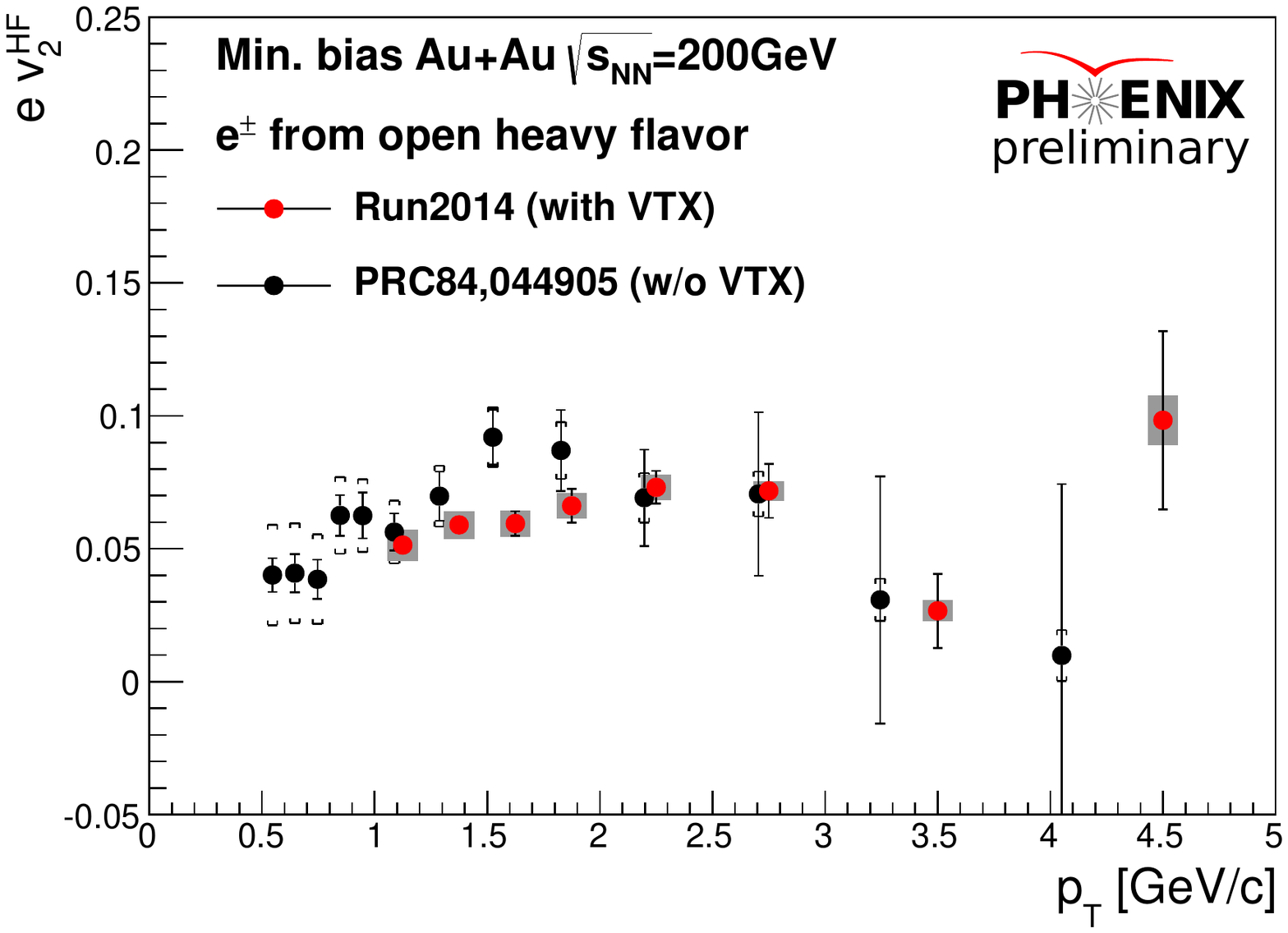}
\caption{Heavy flavor semi-leptonic decayed single electron v$_{2}$ measured with the PHENIX mid-rapidity VTX detector versus $p_{T}$ (red) in 200 GeV Au+Au collisions from the 2014 run in comparison with the previous published results (black) \cite{hfe_v2}.}
\label{fig:hfe_v2}
\end{figure}

Quarks and gluons loss energy when pass through the QGP. Heavy quarks are expected to loss less energy compared to light quarks due to their large masses ($m_{c,b} >> \Lambda_{QCD}$). Due to the mass dependent energy loss, charm semileptonic decayed single electron $R_{AA}$ is more suppressed than bottom semileptonic decayed single electron $R_{AA}$ measured by PHENIX within $3<p_{T}<4$ GeV$/c$ and $|y|<0.35$ kinematic region in 200 GeV Au+Au collisions \cite{hfe_raa}. Significant inclusive heavy flavor decayed single electron v$_{2}$ has been measured with the similar kinematic region in 200 GeV Au+Au collisions \cite{hfe_v2}. High statistic 2014 200 GeV Au+Au data allow us to study the charm and bottom quark thermal properties in QGP by separating the charm and bottom contributions to open heavy flavor production. Figure \ref{fig:hfe_v2} shows the preliminary result of inclusive heavy flavor decayed single electron v$_{2}$ using the mid-rapidity silicon vertex detector (VTX) in 2014 Au+Au data, which is consistent with the previous published results \cite{hfe_v2}. 

\begin{figure}[!ht]
\centering
	\includegraphics[width=0.45\textwidth]{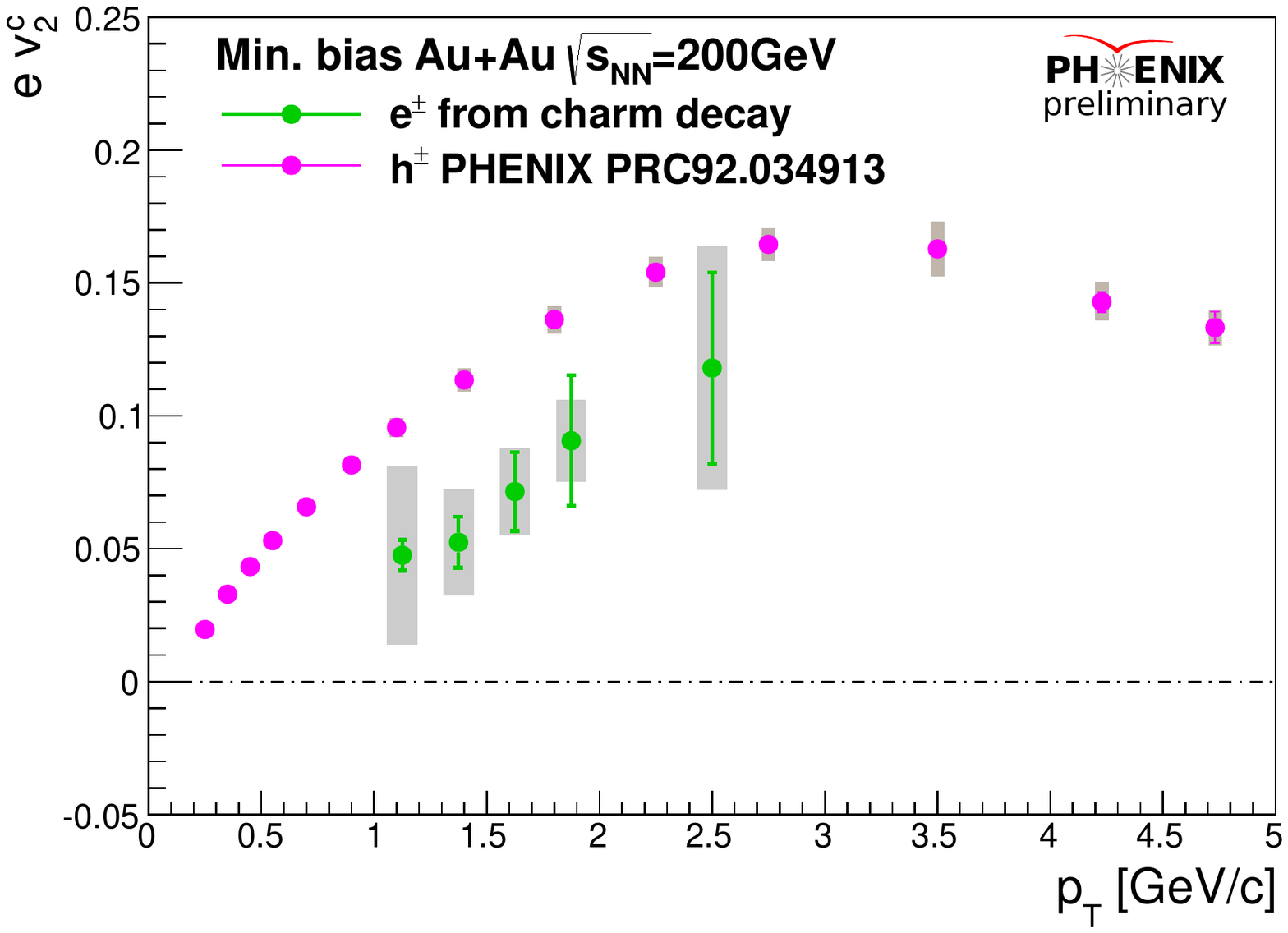}
        \includegraphics[width=0.45\textwidth]{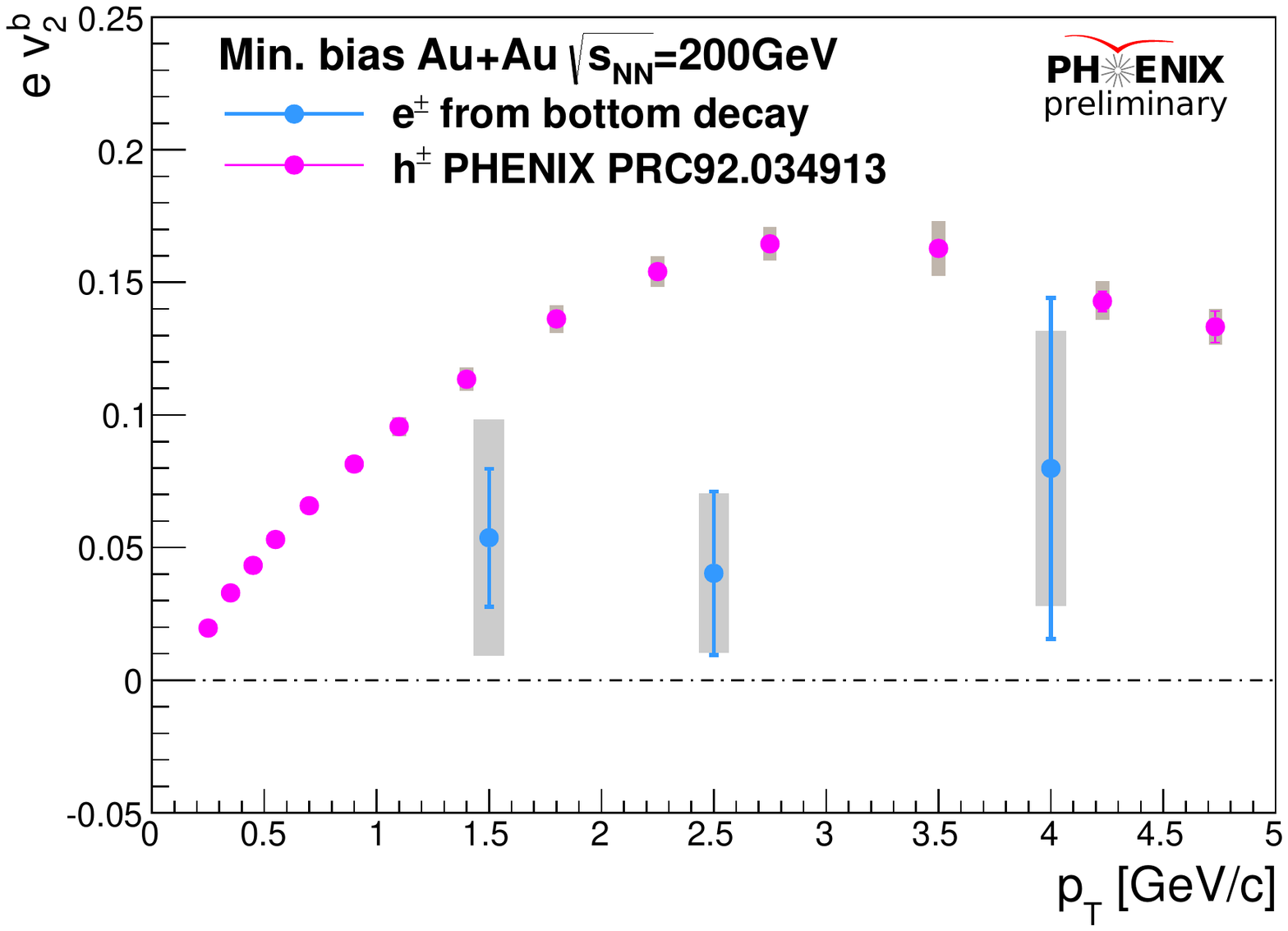}
\caption{Preliminary results of the $p_{T}$ dependent charm decayed single electron v$_{2}$ (green points in the left panel) and bottom decayed single electron v$_{2}$ (blue points in the right panel) within $|y|<0.35$ rapidity region in 200 GeV Au+Au collisions. The previous published charged hadron v$_{2}$ (shown in magenta) measured in the same rapidity region is presented for reference.}
\label{fig:cb_v2}
\end{figure}

\begin{figure}[!ht]
\centering
	\includegraphics[width=0.45\textwidth]{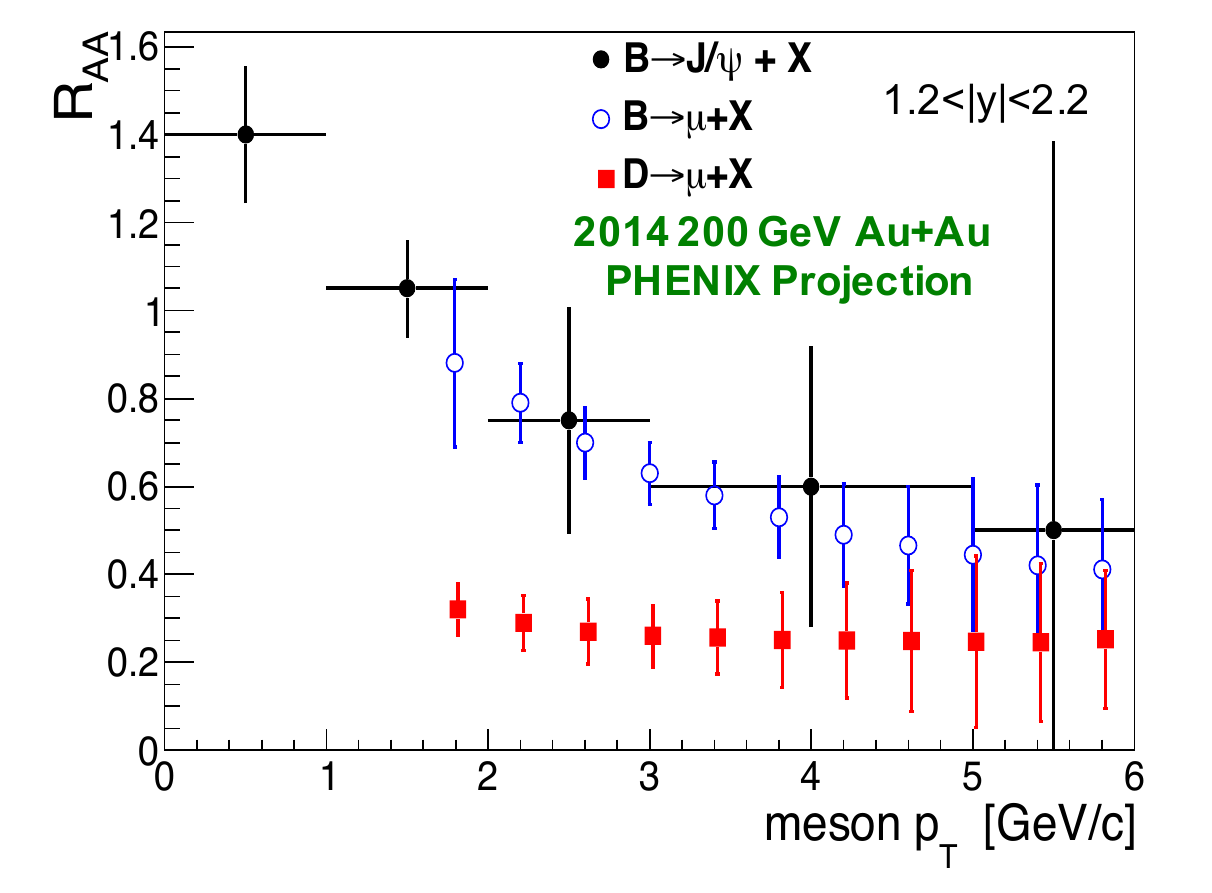}
\caption{Projection of meson $p_{T}$ dependent B $\rightarrow J/\psi$ and charm/bottom semi-leptonic decayed single muon $R_{AA}$ within $1.2<|y|<2.2$ measured by the PHENIX FVTX detector in 2014 200 GeV Au+Au data.}
\label{fig:fvtx_proj}
\end{figure}

The VTX detector can precisely measure the Distance of Closest Approach (DCA) of single electrons which is proportional to the decay lengths of their parents. The DCA analysis can separate the charm semileptonic decayed electrons from the bottom semileptonic decayed electrons and determine their relative fractions. Due to differences in decay lengths between the charm and bottom hadrons, the single electron events are divided into charm-enriched sample ($|DCA| <$ 200 $\mu$m) and bottom-enriched sample (300 $< |DCA| <$ 1000 $\mu$m). By measuring the inclusive heavy flavor decayed electron v$_{2}$ and applying the determined charm, bottom and background fractions in the charm-enriched and bottom-enriched samples separately, the charm and bottom decayed single electron v$_{2}$ can be extracted simultaneously. 

Figure \ref{fig:cb_v2} shows the $p_{T}$ dependent charm decayed single electron v$_{2}$ (green points) and bottom decayed single electron v$_{2}$ (blue points) in comparison with previous published charged hadron v$_{2}$ results \cite{had_AA_v2} in 200 GeV Au+Au collisions. Smaller magnitude of the charm decayed electron v$_{2}$ than charged hadron v$_{2}$ might be caused by the smearing effects from the charm decay kinematics. Significant charm decayed electron v$_{2}$ indicates that charm quarks might flow like light quarks inside the QGP. The first measurement of bottom decayed electron v$_{2}$ indicates that bottom quarks might flow inside the QGP but with smaller magnitude than charm quarks. The forward silicon vertex detector (FVTX) at PHENIX provides precise verte/tracking determination in the forward rapidity region ($1.2<|y|<2.2$) and help release the first B $\rightarrow J/\psi$ measurements at RHIC \cite{bjpsi_pp,bjpsi_CuAu}. Ongoing charm and bottom separated single muon and B $\rightarrow J/\psi$ analysis with the FVTX in high statistics $p$+$p$ and Au+Au data (see projection in Figure \ref{fig:fvtx_proj}) will extend the kinematic region of the open heavy flavor studies at RHIC and provide further information about the heavy quark hot nuclear modification and azimuthal anisotropy within the QGP.

\section{Summary and Outlook}
PHENIX experiment has obtained several new heavy flavor measurements at mid-rapidities and forward/backward rapidities in different collisions systems. Both initial and final state effects play an important role in the inclusive $J/\psi$ production in asymmetric small systems. Heavy quarks especially charm quarks are found not only flow in large systems such as Au+Au collisions but also flow in small systems such as d+Au collisions. Large statistics data of $p$+$p$, asymmetric nuclear and heavy ion collisions collected at PHENIX allow more precise heavy flavor measurements in different kinematic regions to explore the energy loss mechanism and thermal properties of the QGP.


\end{document}